\documentclass[12pt,a4paper]{article}

\title{On asymptotic properties of some complex Lorenz-like systems}
\author{S. Panchev $^{*}$ and N. K. Vitanov $^{**}$}
\date{}
\begin{document}
\maketitle
\begin{abstract}
The classical Lorenz lowest order system of three nonlinear
ordinary differential equations, capable of producing chaotic
solutions, has been generalized by various authors in two main
directions: (i) for number of equations larger than three
(Curry1978) and (ii) for the case of complex variables and
parameters. Problems of laser physics and geophysical fluid
dynamics (baroclinic instability, geodynamic theory, etc. - see
the references) can be related to this second aspect of
generalization. In this paper we study the asymptotic properties
of some complex Lorenz systems, keeping in the mind the physical
basis of the model mathematical equations.
\end{abstract}
\section{Introduction}
The behaviour of the classical Lorenz system of equations (Lorenz 1963)
\begin{eqnarray}\label{lorenz}
\frac{dX}{dt}= - \sigma X + \sigma Y, \nonumber \\
\frac{dY}{dt}= r X - Y - XZ,  \nonumber \\
\frac{dZ}{dt}= -bZ + XY,
\end{eqnarray}
where $\sigma >0, r >0, b>0$, is well known (Shimizu and Morioka 1978,
Yorke and Yorke 1979, Franceschini 1980, Sparrow 1982, McGuinnes 1983,
Schmutz and Rueff 1984). Here
$X(t), Y(t), Z(t)$ are real functions of the time $t$ and $(\sigma, r,b)$
are real parameters. The goal of this study is to ascertain to what
extent some fundamental properties of the system (\ref{lorenz}) are
presented in its complex analogs. For comparison we choose the following
properties of the system of equations (\ref{lorenz}):
\begin{enumerate}
\item
The fixed points of (\ref{lorenz})
\begin{equation}\label{fp}
(\overline{X},\overline{Y},\overline{Z})=(0,0,0), \hskip.25cm r>0,
\end{equation}
\begin{equation}\label{fixpoint}
(\overline{X} = \overline{Y} = \pm \sqrt{b \overline{Z}}, \overline{Z}=
r-1), \hskip.25cm r>1.
\end{equation}
\item The conditions for stability of the second fixed fixed point
in (\ref{fixpoint}) are
\begin{equation}\label{stability1}
1 < r < r_{H} = \frac{\sigma (\sigma + b+3)}{\sigma - b - 1}, b+1 < \sigma <
\infty .
\end{equation}
At $r >r_{h}$ a chaos occurs.
\item
If $\sigma \to \infty$ then $r_{H} \to \infty$ and the above fixed point
remains stable for arbitrary $r>1$. In this case
\begin{eqnarray}\label{lorenz1}
Y(t) = X(t) ,\nonumber \\
X^{2} = \frac{dZ}{dt} + bZ, \nonumber \\
\frac{d^{2} Z}{dt^{2}} + (b - 2f + 2Z) \frac{dZ}{dt} - 2bZ(f-Z) =0,
\end{eqnarray}
where $f=r-1$, so that chaos is impossible.
\item
If $b=2 \sigma$ and $t \to \infty$, then
\begin{eqnarray} \label{lorenz2}
Y(t) = X \frac{1}{\sigma} \frac{dX}{dt} \nonumber, \\
Z(t) = \frac{1}{2 \sigma} X^{2} \nonumber, \\
\frac{d^{2} X}{dt^{2}} + (\sigma +1) \frac{dX}{dt} + \sigma (1-r) X +
\frac{1}{2} X^{3} =0,
\end{eqnarray}
so that again chaos is impossible. The single second order
autonomous equations in (\ref{lorenz1}) and (\ref{lorenz2}) are of
known types (Lienard and Duffing respectively - Perko 1996).
Their derivation can be seen below.
\end{enumerate}
\par
The Lorenz system (\ref{lorenz})  is a very simplified model
of thermal convection in fluids (the Rayleigh-Benard problem,
see Lorenz 1963 or Tritton 1988).
Later on, problems of laser physics (see Ning and Haken, 1990 and the
references therein) and of so called baroclinic instability of the
geophysical flows (in the atmosphere or in the ocean) unexpectedly resulted in a system
of the same structure like (\ref{lorenz}) but in the complex domain
(Fowler et al. 1983, Rauh et al. 1996)
\begin{eqnarray}\label{complex}
\frac{dX}{dt}= - \sigma X + \sigma Y, \nonumber \\
\frac{dY}{dt}= \hat{r} X - aY -XZ, \nonumber \\
\frac{dZ}{dt}= - bZ + \frac{1}{2} (X^{*} Y + X Y^{*}).
\end{eqnarray}
where $i= \sqrt{-1}$, $\hat{r} = r_{1} + i r_{2}, a = 1- i e,
(X,Y) = (X_{1},Y_{1}) + i(X_{2},Y_{2})$, $(\sigma, b, e, Z)$ are real
quantities and $^{*}$ denotes complex conjugate. Another similar system
originated from geophysics will be introduced in section 4.
\par
In real variables the system (\ref{complex}) is equivalent to the
following fifth order system of ordinary differential equations
\begin{eqnarray}\label{complex2}
\frac{d X_{1}}{dt} = -\sigma X_{1} + \sigma Y_{1}, \nonumber \\
\frac{d X_{2}}{dt} = -\sigma X_{2} + \sigma Y_{2}, \nonumber \\
\frac{dY_{1}}{dt}= r_{1} X_{1} -Y_{1} -r_{2} X_{2} - e Y_{2} - X_{1} Z,
\nonumber \\
\frac{dY_{2}}{dt}= r_{1} X_{2} -Y_{2} -r_{2} X_{1} - e Y_{1} - X_{2} Z,
\nonumber \\
\frac{dZ}{dt}= - Z + X_{1} Y_{1} + X_{2} Y_{2}.
\end{eqnarray}
In what follows we show that in some particular cases the above
system simplifies and can be integrated even analytically.
\section{Particular cases of the system (\ref{complex})}
\subsection{Case $\sigma \to \infty$}
Let $\sigma \to \infty$, i.e., $\epsilon = 1/\sigma \to 0$. Then the
first equation of (\ref{complex}) implies
\begin{equation}\label{eq1}
Y(t) = X(t),
\end{equation}
so that (\ref{complex}) reduces to
\begin{equation}\label{eq2}
\frac{dX}{dt} = (\hat{r} - a - Z) X,
\end{equation}
\begin{equation}\label{eq3}
\frac{dZ}{dt}= -bZ + \mid X \mid^{2},
\end{equation}
where $\mid X \mid^{2} = X X^{*} = X_{1}^{2} + X_{2}^{2} = A^{2} >0$.
In real variables the above equations are equivalent to a three-dimensional
system of nonlinear differential equations
\begin{eqnarray} \label{eq4}
\frac{dX_{1}}{dt}= (f_{1} -Z) X_{1} - \nu X_{2}, \nonumber \\
\frac{dX_{2}}{dt}=(f_{1} -Z) X_{2} + \nu X_{1}, \nonumber \\
\frac{dZ}{dt} = -bZ + X_{1}^{2} + X_{2}^{2},
\end{eqnarray}
where $f_{1}=r_{1}-1$, $\nu = e +r_{2}$. From the first two equations
of (\ref{eq4}) we obtain easily
\begin{equation}\label{eq5}
\frac{d}{dt}A^{2} = 2 (f_{1}-Z) A^{2},
\end{equation}
Eliminating $A^{2}$ from (\ref{eq5}) and (\ref{eq3}) we obtain a
single equation of known (Lienard) type (Perko 1996)
\begin{equation}\label{lienard}
\frac{d^{2} Z}{dt^{2}}+(b-2f_{1} +2Z) \frac{dZ}{dt} - 2bZ(f_{1}-Z) =0,
\end{equation}
In addition by eliminating either $X_{1}$ or $X_{2}$ from the
first two equations of (\ref{eq4}) we obtain the second order
equation of kind forced harmonic oscillator
\begin{equation}\label{eq6}
\frac{d^{2}F}{dt^{2}}-2(f_{1} -Z) \frac{dF}{dt} + \left[ \nu^{2}+
(f_{1}-Z)^{2} + \frac{dZ}{dt} \right] F = 0,
\end{equation}
where $F$ is equal to $X_{1}$ or to $X_{2}$. The equations
(\ref{lienard}) and (\ref{eq6}) form a decoupled system. Having
determined (analytically or numerically) $Z(t)$ from
(\ref{lienard}) then we can obtain $A$ by $A^{2}= dZ/dt + bZ$ and
$X_{1,2}$ can be determined by (\ref{eq6}). For instance
\begin{equation}\label{eqx2}
\overline{Z}=f_{1} = r_{1}-1,
\overline{A}^{2} = bf_{1}, \hskip.5cm (r_{1}>1),
\end{equation}
are stationary ($dZ/dt =0$) solutions
(fixed points) of (\ref{eq5}) and (\ref{eq3}) for which (\ref{eq6}) reduces
to the simplest (linear) harmonic oscillator $d^{2} F /dt^{2} + \nu^{2}F =0$.
 Hence $\overline{A} = \sqrt{b f_{1}}$ and
\begin{equation}\label{eq7}
\overline{X_{1}}(t) = \overline{A} \cos (\nu t), \hskip.25cm
\overline{X_{2}}(t) = \overline{A} \sin (\nu t)
\end{equation}
Thus (\ref{eq7}) is a simple limit cycle solution, which is a result of a Hopf
bifurcation after loss of stability at $r_{1}=1$ of the trivial solution
$\overline{Z} = \mid \overline{X} \mid = 0$ of (\ref{eq3}) and
(\ref{eq5}). Standard stability analysis of the above fixed point shows that
it is stable for entire range $r_{1}>1$ so that (\ref{eq7}) is a
stable cycle for all $r_{1}>1$.
\par
In the general case the Lienard equation (\ref{lienard}) can not be
solved analytically. For its numerical integration a more appropriate
form is
\begin{eqnarray} \label{eqx}
\frac{dZ}{dt}= U - (b-2f_{1}) Z - Z^{2}, \nonumber \\
\frac{dU}{dt}=2 b Z (f_{1}-Z).
\end{eqnarray}
If $b=2f_{1}$ both equations (\ref{lienard}) and (\ref{eqx}) simplify.
\par
It is of interest to study the behavior of $Z(t)$ and $A^{2}(t)$
in (\ref{eq3}) and (\ref{eq5}) far from the initial instant, i.e.,
in the posttransient period.  The general solution of (\ref{eq3})
with respect to $Z(t)$ can be written as
\begin{equation}\label{eq8}
Z(t)=Z_{0} \exp (-bt) + \int_{0}^{t} d \tau \exp(-b \tau) \mid X(t-\tau). \mid^{2}
\end{equation}
Hence at $t >> \tau_{b} = 1/b$ (theoretically $t \to \infty$)
\begin{equation}\label{eq9}
Z( t \to \infty) = Z_{\infty}(t) = \int_{0}^{\infty}  d \tau \exp(- b \tau)
\mid X(t- \tau) \mid^{2} .
\end{equation}
The substitution of (\ref{eq5}) for $Z(t)$ yields the
integrodifferential equation
\begin{equation}\label{eq10}
\frac{d}{dt} \mid X^{2} \mid = 2 \left[ f_{1} - \int_{0}^{\infty} d \tau
 \exp(- b \tau) \mid X(t- \tau) \mid^{2} \right] \mid X \mid^{2}.
\end{equation}
valid at $t \to \infty$. The integral term can be interpreted as memory of the
process for the past. If in addition $b >>1$ (in the limit $b \to \infty$)
, i.e., in the case of short memory, the expression (\ref{eq9})
is reduced to
\begin{equation}\label{eq11}
Z_{\infty} (t) = \frac{1}{b} \mid X(t) \mid^{2} = \frac{1}{b} A^{2} (t),
\end{equation}
so that from (\ref{eq10})
\begin{equation}\label{eq12}
\frac{d}{dt} A^{2} = 2 f_{1} A^{2} - \frac{2}{b} A^{4}, \hskip.5cm
t \to \infty .
\end{equation}
However (\ref{eq12}) is exactly the Landau equation in turbulence
theory (Landau and Lifshitz, 1986). Its integration is an easy task.
\begin{equation}\label{eq13}
A^{2}(t) = A_{0}^{2} b f_{1} \left[ A_{0}^{2} + (b f_{1} - A_{0}^{2})
\exp(- 2 f_{1}(t-t_{0}))\right]^{-1}
\end{equation}
where $A_{0}^{2}=[A(t_{0})]^{2}$. Therefore, independently of $A_{0}$,
 $A^{2} (\infty) = b f_{1} = \overline{A}^{2}$, which means that the fixed points
(\ref{eqx2}) and the corresponding cycle (\ref{eq7}) are globally stable.
\subsection{Case of finite $\sigma$ ($\sigma < \infty$)}
Let us now assume finite $\sigma$ ($\sigma < \infty$). Then from
the $X$, $Z$ equations (\ref{eq4}) we derive a new one
\begin{equation}\label{eq14}
\frac{dZ}{dt} + bZ = \mid X \mid ^{2} + \frac{1}{2 \sigma} \frac{d}{dt}
\mid X \mid^{2},
\end{equation}
which can be rewritten as
\begin{equation}\label{eq15}
\frac{dM}{dt} + bM = \left( \frac{b}{2 \sigma} -1 \right) \mid X \mid^{2},
\end{equation}
or alternatively
\begin{equation}\label{eq16}
\frac{dM}{dt} + 2 \sigma M = (b - 2 \sigma) Z(t),
\end{equation}
where
\begin{equation}\label{eq17}
M(t) = \frac{1}{2 \sigma} \mid X \mid^{2} -Z.
\end{equation}
Hence in view of (\ref{eq3}) and (\ref{eq9})
\begin{equation}\label{eq18}
M_{\infty}(t) = M( t \to \infty) = \left( \frac{b}{2 \sigma} - 1\right)
\int_{0}^{\infty} d \tau \exp(-b \tau) \mid X (t- \tau) \mid^{2},
\end{equation}
and
\begin{equation}\label{eq19}
Z_{\infty} (t) = \frac{1}{2 \sigma} \mid X(t) \mid^{2} - M_{\infty} (t),
\end{equation}
so that $M_{\infty}=0$ at $b=2 \sigma$.  Then the first two equations
(\ref{eq4})  combined with  (\ref{eq19})  lead to a single equation
\begin{equation}\label{eq20}
\frac{d^{2} X}{dt^{2}}+ (\sigma + a) \frac{dX}{dt}+ \sigma ( a - \hat{r})
X + \frac{1}{2} \mid X \mid^{2} X =0,
\end{equation}
valid at $t \to \infty$ and $b=2 \sigma$. Obviously, this is a generalized
autonomous Duffing type equation in the complex domain. In real variables it
is equivalent to two coupled equations for $X_{1}(t)$ and $X_{2}(t)$ of the
same kind
\begin{eqnarray}\label{eqx3}
\frac{d^{2} X_{1}}{d t^{2}} + (\sigma +1) \frac{d X_{1}}{dt} +
e \frac{d X_{2}}{dt} - \sigma f_{1} X_{1} + \sigma \nu  X_{2} +
\frac{1}{2} \mid X \mid^{2} X_{1}=0, \nonumber \\
\frac{d^{2} X_{2}}{d t^{2}} + (\sigma +1) \frac{d X_{2}}{dt} +
e \frac{d X_{1}}{dt} - \sigma f_{1} X_{2} + \sigma \nu  X_{1} +
\frac{1}{2} \mid X \mid^{2} X_{2}=0,
\end{eqnarray}
where $\mid X \mid^{2} = X_{1}^{2} + X_{2}^{2}$, $f_{1} = r_{1} -1$,
$\nu = e + r_{2}$ and $X = X_{1} + i X_{2}$.
\par
Since $a- \hat{r} = -f_{1} - i \nu$, equation (\ref{eq20}) has
only one stationary solution (fixed point): $\overline{X}=0$. No
other such solutions occur unless $\nu =0$, known as "laser case"
(Ning and Haken 1990).  Then
\begin{equation}\label{eq21}
\mid \overline{X} \mid^{2} = 2 \sigma (r_{1}-1), \hskip.5cm r_{1}>1,
\end{equation}
and stationary oscillations in the lasers can exist. However, such constraint
($\nu =0$) is not generally required for the baroclinic instability.
In the latter case $\nu \ne 0$.
\begin{equation}\label{eq22}
X(t)=A \exp(i \omega t), \omega = \frac{\sigma \nu}{\sigma +1},
A^{2}= 2 \sigma \left[ f_{1} - \nu \frac{e-\sigma r_{2}}{(\sigma +1)^{2}}\right],
\end{equation}
is an exact limit cycle solution of the nonlinear equation (\ref{eq20})
respectively (\ref{eqx3}), provided the term in the bracket is positive,
i.e., $r_{1} > r_{1c} = 1 + \nu (e -\sigma r_{2})/(\sigma +1)^{2}$.
\par
If $b \ne 2 \sigma$, the memory term $\sigma M_{\infty} (t)$ will
stand in (\ref{eq20}) instead of zero. This changes the expression for
the amplitude $A^{2}$ only: $2 \sigma$ is replaced by $b$ ($b< 
2 \sigma$ or $b> 2 \sigma$),
thus decreasing (increasing) $A^{2}$ compared to (\ref{eq22}).
\par
For the case of finite $\sigma$ but $ b \to 0$ we can use the
the alternative equation (\ref{eq16})
\begin{eqnarray}
\frac{d M}{dt} + 2 \sigma M = - 2 \sigma Z(t).
\nonumber
\end{eqnarray}
Hence similarly to (\ref{eq18}) and (\ref{eq19})
\begin{eqnarray}
M_{\infty}(t)= - 2 \sigma \int_{0}^{\infty} d \tau \hskip.05cm
\exp(-2 \sigma \tau) Z_{\infty} (t- \tau),
\nonumber
\end{eqnarray}
and
\begin{eqnarray}
\frac{1}{2 \sigma} \mid X_{\infty} (t) \mid^{2} = Z_{\infty} (t) +
M_{\infty} (t) .
\nonumber
\end{eqnarray}
Contrary to the previous case (\ref{eq9}) the memory effect now is
carried out by the phase variable $Z(t)$.
\par
Finally, we note that equation (\ref{lienard}) is identical to the
last equation (\ref{lorenz1}) and  equation (\ref{eq20})
degenerates into equation (\ref{lorenz2}) in the real case ($e
=r_{2}=0, X_{2}=0, X_{1}=X, r_{1}=r$).
\section{The infinite $Z-$components version of (\ref{complex})}
The paper by Booty et al. (1982) contains results for the system
\begin{eqnarray}\label{complex3}
\frac{d X}{dt} = - \sigma X + \sigma Y, \nonumber \\
\frac{d Y}{dt} = (\hat{r} - \sum_{n=1}^{\infty} Z_{n} ) X - a Y, \nonumber \\
\frac{d Z_{n}}{dt} = - b_{n} Z_{n} + \frac{1}{2} c_{n} (X^{*} Y + X Y^{*}),
\end{eqnarray}
where $b_{n}, c_{n} >0$, $n = 1,2 \dots, \infty$, $(\sigma , \hat{r}, a)$
are the same parameters as in (\ref{complex}). However, unlike
(\ref{complex}), the physical basis of the above system is the
baroclinic instability in the geophysical fluids only. Here we are
interested in (\ref{complex3})  from mathematical point of view.
\par
First of all if
\begin{equation}\label{eq23}
b_{n}=b= {\rm const}, \hskip.25cm \sum_{n=1}^{\infty} c_{n} = c < \infty,
\end{equation}
the system (\ref{complex3}) becomes identical to (\ref{complex}) with
$Z= \sum_{n=1}^{\infty} Z_{n}$. We now apply the approach from the
section 2.2. The equation analogous to (\ref{eq15}) here reads
\begin{eqnarray}
\frac{d M_{n}}{dt} + b_{n} M_{n} = c_{n} \left(\frac{b_{n}}{2 \sigma}
- 1 \right) \mid X \mid^{2},
\nonumber
\end{eqnarray}
where
\begin{eqnarray}
M_{n} = \frac{c_{n}}{2 \sigma} \mid X \mid^{2} - Z_{n}.
\nonumber
\end{eqnarray}
Hence, at $b_{n} = 2 \sigma$ and $t \to \infty$ we have $M_{m} \to 0$,
so that
\begin{eqnarray}
Z_{n}(t \to \infty) = \frac{c_{n}}{2 \sigma} \mid X \mid^{2}
\nonumber
\end{eqnarray}
Further derivations follow those of section 2.2 with final result
equation (\ref{eq20}) with the last term on the left hand side
multiplied by $c= \sum c_{n}$. Its contributions depends on the
value of $c$. If one adopts the model speculations from Booty et
al. (1982) and assume $c_{n}=n^{-p}$, $p>1$, then
\begin{equation}\label{eq24}
c = \sum_{n=1}^{\infty} n^{-p} = \zeta(p), \hskip.25cm p>1
\end{equation}
where $\zeta(p)$ is the Rieman $\zeta$-function. We note that
Curry (1978) and Curry et al. (1984) discussed system similar to
(\ref{complex3}) for the system (\ref{lorenz})
\section{Another complex Lorenz system}
Of the several complex Lorenz-like systems known in the nonlinear
geodynamo theories (Weiss et al. 1984, Jones et al. 1985, Roberts
and Glatzmaier, 2000), the most interesting appears to be the
following
\begin{eqnarray}\label{complex4}
\frac{dA}{dt}= \sqrt{\sigma} (\sigma +1) DB - \sigma A, \nonumber \\
\frac{dB}{dt} = i A - \frac{1}{2} i C A^{*} - B, \nonumber \\
\frac{dC}{dt} = -mC - i A B,
\end{eqnarray}
where unlike (\ref{complex}) all geodynamo characteristics (phase
variables) $A(t)$, $B(t)$, $C(t)$ are complex functions of the
kind $K(t) = K_{1}(t) + i K_{2}(t)$, $K=A,B,C$, whereas $D,
\sigma,m$ are real physical parameters and in addition $\sigma
\propto m \propto 1$. Thus in the real domain (\ref{complex4}) is
a sixth order nonlinear dynamical system. According to Weiss et
al. (1984) it exhibits much richer and more fascinating variety of
behaviour than other similar systems discussed  in the cited
paper. Below we shall concentrate our attention on it.
\par
Following Jones et al. (1985) we set
\begin{eqnarray}
A = \sqrt{2} X, B=(i \sqrt{2} /R) Y, C=2Z/R, R= i(\sigma+1)D /\sqrt{\sigma} .
\nonumber
\end{eqnarray}
Then (\ref{complex4}) takes the form
\begin{eqnarray}\label{complex5}
\frac{dX}{dt}= - \sigma X + \sigma Y, \nonumber \\
\frac{dY}{dt}= RX - Y- X^{*} Z, \nonumber \\
\frac{dZ}{dt}= - m Z + XY,
\end{eqnarray}
which closely resembles the systems (\ref{lorenz}) and (\ref{complex}).
This fact implies that the subsequent results if not identical, will be
similar to those corresponding to (\ref{lorenz}) and (\ref{complex}).
For instance, at $m= 2 \sigma$ and $t >> 1/m$ $(t \to \infty)$, the equation
(\ref{eq20}) now reads
\begin{equation}\label{eq25}
\frac{d^{2} X}{dt^{2}} + (\sigma +1) \frac{dX}{dt} + \sigma (1-R) X
+ \frac{1}{2} \mid X \mid^{2} X =0.
\end{equation}
Unlike (\ref{eq20}), $\overline{X} =0$ is the only stationary
solution of (\ref{eq25}). On the other hand similar to (\ref{eq20})
and (\ref{eq22}) now
\begin{equation}\label{eq26}
X(t)= \delta \exp(i \omega t), \omega = \sqrt{\sigma} D,
\delta^{2} = 2 \delta (D^{2}-1), (D>1),
\end{equation}
is an exact periodic solution of (\ref{eq25}). In details the
stability and the bifurcation properties of the fixed point
$\overline{X}=0$ and of (\ref{eq26}) are described in Jones et al.
(1985).
\par
The physical constraints $\sigma \propto m \propto 1$ make the
previous cases ($\sigma \to \infty$ and /or $m \to 0$) invalid
for (\ref{complex5}).
\section{Summary and conclusion}
The main results of this study concern the complex Lorenz-like systems
(\ref{complex}) and its infinite $Z$-component generalization
(\ref{complex3}) as well as the system (\ref{complex4}), arising
in the laser physics and in geophysical fluid dynamics (baroclinic instability
theory, geodynamo models, etc.). Some of them recover the results obtained
by previous authors. Here we were interested mainly in the asymptotic
solutions of the corresponding equations:
\begin{itemize}
\item
At $\sigma \to \infty$ two decoupled equations (\ref{lienard}), (\ref{eq6})
of known type are derived from the reduced system (\ref{eq1}) - (\ref{eq3})
and the solutions (\ref{eqx2}), (\ref{eq7}) are discussed.
\item
At $t \to \infty$ a single equation (\ref{eq10}) with integral (memory)
term is derived. In the case of short memory ($1/b \to 0$), equations
(\ref{eq11}) - (\ref{eq13}) are in force.
\item
At finite $\sigma$ ($\sigma < \infty$) , the generalized Duffing type
equation (\ref{eq20}) with exact limit cycle solution (\ref{eq22}) is derived.
Almost identical to (\ref{eq20}) equation (\ref{eq25}) is derived for the
nonlinear geodynamo model (\ref{complex4}). Unlike (\ref{eq20}),
equation (\ref{eq25}) has a zero fixed point only.
\item
Under the same assumptions, the generalized system (\ref{complex3})
is reduced to the standard form (\ref{complex}) with $Z=\sum_{n=1}^{\infty} Z_{n}$
\end{itemize}
Our study open possibility for further numerical  and analytical
investigation of the properties of obtained equations such as
investigation of stability properties of attractors, bifurcation
diagram and calculation of characteristic quantities such as
generalized dimensions . This will be a subject of future
research.

\section{Acknowledgments}
The research of N. K. V. was  supported by the contract MM
1201/02 with the National Science Fund of the Ministry of
Education and Science of Republic of Bulgaria. \vskip1cm
\begin{flushleft}
{\Large \sc \bf References}
\end{flushleft}
\vskip.5cm
\begin{description}
\item[]
Booty M., Gibbon J. D., and Fowler A. C. (1982): A study of the
    effect of mode truncation on an exact periodic solution of an infinite
    set of Lorenz equation. {\sl Phys. Lett} {\bf 87} A, 6, 261-266.
\item[]
Curry J. C. (1978): A generalized Lorenz system. {\sl Commun. Math.
    Physics} {\bf 60}, 193-204.
\item[]
Curry J. H., Herring J. R., Loncaric J., Orszag S. A. (1984): Order and
    disorder in two- and three-dimensional Benard convection. {\sl J.
    Fluid Mech.} {\bf 147}, 1-36.
\item[]
Fowler A. C., Gibbon J. D., McGuinness M. J. (1983): The real and
    complex Lorenz equations and their relevance to physical systems.
    {\sl Physica} D {\bf 7}, 126-134.
\item[]
Franceschini V. A. (1980). A Feigenbaum sequence of bifurcations in the
Lorenz model. {\sl J. Stat. Phys.} {\sl 22}, 397-406.
\item[]
Jones C. A., Weiss N. D., and Cataeno F. (1995): Nonlinear dynamos:
    A complex generalization of the Lorenz equations. {\sl Physica} D
    {\bf 14}, 161-176.
\item[]
Landau L. D. and Lifshitz E. M. (1986): Hydrodynamics. Nauka, Moscow
(in Russian).
\item[]
Lorenz  E. N. (1963). Deterministic nonperiodic flow. {\sl J. Atmos. Sci.}
    {\bf 20}, 130-141.
\item[]
McGuiness M. J. (1983). The fractal dimension of the Lorenz attractor.
{\sl Phys. Lett. A} {\bf 99}, 5-9.
\item[]
Ning C. -Z., Haken H. (1990). Detuned lasers and complex Lorenz equations -
subcritical and supercritical Hopf bifurcations. {\sl Phys. Rev.} A
{\bf 41}, 3827-3837.
\item[]
Perko L. (1996). Differential equations and dynamical systems.
    Springer, New York, 519.
\item[]
Rauh A., Hannibal L., and Abraham N. D. (1996). Global stability of the
complex Lorenz equations. {\sl Physuca } D {\bf 99}, 45-58.
\item[]
Roberts P. H. and Glazmaier G. A. (2000): Geodynamo theory and simulations.
    {\sl Rev. Mod. Phys} {\bf 72}, 4, 1083-1123.
\item[]
Schmutz M., and Rueff M. (1984). Bifurcation schemes of the Lorenz model.
    {\sl Physica D} {\bf 11}, 167-178.
\item[]
Shimizu T., and Morioka N. (1978):
    Transient behaviour in periodic regions of the Lorenz model.
    {\sl Phys. Lett. A} {\bf 69}, 148-150.
\item[]
Sparrow C. T. (1982). The Lorenz equations: Bifurcations, chaos and
    strange attractors. Springer, Berlin, 270.
\item[]
Tritton D. J. (1988). {\sl Physical fluid dynamics}. Clarendon Press,
Oxford.
\item[]
Weiss N. D., Cattaeno F., and Jones C. A. (1984):
    Periodic and apperiodic dynamo wakes. {\sl Geophys. Astrophys.
    Fluid Dyn.} {\bf 30}, 305-341.
\item[]
Yorke J. A., and Yorke E. D. (1979):
    The transition to sustained chaotic behaviour in the Lorenz model.
    {\sl J. Stat. Phys.} {\bf 21}, 263-277.
\end{description}
\vskip2cm
\begin{flushleft}
{\sc $^{*}$ Solar-Terrestial Influences Laboratory \\
Bulgarian Academy of Sciences \\
akad. G. Bonchev Str., Bl. 3 , 1113 Sofia, Bulgaria \\
and Faculty of Physics, "St. Kliment Okhridski" University of Sofia 
1164 Sofia, Bulgaria \\
e-mail}: spanchev@phys.uni-sofia.bg
\end{flushleft}
\begin{flushleft}
{ \sc $^{**}$ Institute of Mechanics \\
Bulgarian Academy of Sciences \\
akad. G. Bonchev Str., Bl. 4, 1113 Sofia, Bulgaria \\
and Max-Planck-Institute for the Physics of Complex Systems, 
Noetnitzer Str. 38, 01187 Dresden, Germany \\
e-mail}: vitanov@imech.imbm.bas.bg \\
\end{flushleft}
\end{document}